# Thermo-mechanical FE model with memory effect for 304L austenitic stainless steel presenting microstructure gradient
*IJFat 45 2012 106-115*

A. Le Pécheur[a,b], F. Curtit[b], M. Clavel[a], J.M. Stephan[b], C. Rey[a], Ph. Bompard[a]

[a]Laboratoire MSSMat, UMR 8579 CNRS, Ecole Centrale Paris, Grande voie des Vignes, 92295 Châtenay-Malabry Cedex France
[b] Département MMC, EDF R&D, Site des Renardières, Route de Sens Ecuelles, 77250 Moret-sur-Loing, France

**Abstract**

The main purpose of this study is to determine, via a three dimensions Finite Element analysis (FE), the stress and strain fields at the inner surface of a tubular specimen submitted to thermo-mechanical fatigue. To investigate the surface finish effect on fatigue behaviour at this inner surface, mechanical tests were carried out on real size tubular specimens under various thermal loadings. X ray measurements, Transmission Electron Microscopy observations and micro-hardness tests performed at and under the inner surface of the specimen before testing, revealed residual internal stresses and a large dislocation microstructure gradient in correlation with hardening gradients due to machining. A memory effect, bound to the pre-hardening gradient, was introduced into an elasto-visco-plastic model in order to determine the stress and strain fields at the inner surface. The temperature evolution on the inner surface of the tubular specimen was first computed via a thermo-elastic model and then used for our thermo-mechanical simulations. Identification of the thermo-mechanical model parameters was based on the experimental stabilized cyclic tension-compression tests performed at 20°C and 300°C. A good agreement was obtained between numerical stabilized traction-compression cycle curves (with and without pre-straining) and experimental ones. This 3 dimensional simulation gave access to the evolution of the axial and tangential internal stresses and local strains during the tests. Numerical results showed: a decreasing of the tangential stress and stabilization after 40 cycles, whereas the axial stress showed weaker decreasing with the number of cycles. The results also pointed out a ratcheting and a slightly non proportional loading at the inner surface. The computed mean stress and strain values of the stabilized cycle being far from the initial ones, they could be used to get the safety margins of standard design related to fatigue, as well as to get accurate loading conditions needed for the use of more advanced fatigue analysis and criteria.

Keywords: 304L stainless steel; thermal fatigue; F.E. modelling.

# 1 Introduction

The 304L stainless steel is a major constituent of residual heat removal circuits of Pressurized Water Reactors (PWR). To assess the risk of thermal fatigue damage [1,2] resulting from the machining of the pipes inner surface (pre-hardening gradient, residual stresses and scratches), experimental tests on tubular specimens (named INTHERPOL tests), were developed by EDF R&D. Other tests facilities were also developed by CEA [3].

Recently, experimental investigations at room temperature on the fatigue behaviour of pre-hardened stainless steel were carried out by Taleb and Hauet [4], Taheri et al [5]. To predict the behaviour of such steels, different models were proposed by Chaboche [6], Taleb and Cailletaud [7] and Said [8].

In this paper, the thermo-mechanical fatigue behaviour of a whole INTHERPOL mock-up was determined at each step of the cycle, via a three dimensional finite element model based on Chaboche's model with memory effect. To validate our simulations, mechanical characterizations (X ray, microstructure observations, micro-hardness tests, temperature gradients at the inner surface of the tube) were performed on the tubular specimens before and after cyclic thermal loading. Computation data were obtained from tensile-compression fatigue tests at 20°C and 300°C, performed on specimens cut out in the initial material, either pre-hardened or not. Involving high temperatures, surface thermal exchanges and cyclic plasticity, such experiments did not enable access to strain fields which evolution during the whole tests had to be computed, as well as the internal stresses. The long term purpose of this experimental and numerical study was to determine the effect of the surface finishing of tubes, on fatigue damage sensitivity, i.e. initiation of fatigue cracks networks. As the scratches depth and width were comparable to grain size, a further fatigue damage study was performed at the grain scale via a polycrystal modelling (submitted).

The first part of this work (section 2) deals with the material characterization of 304L stainless steel, emphasizing the specificities of the surface roughness linked with a strong hardening gradient. Micro-hardness measurements and TEM observations are intensively used to characterize this gradient. The second part (section 3) deals with the macroscopic modelling of INTHERPOL tests and the simulation of the components of the local stress and strain tensors due to thermal cycling. The paper is closed by a discussion and a conclusion (section 4).

# 2 Experimental thermal fatigue tests and characterization of 304L austenitic stainless steel

2.1 Thermal fatigue tests

INTHERPOL thermal fatigue tests were carried out on five 304L austenitic stainless steel tubular specimens. The INTHERPOL facility was designed to apply thermal cycles controlled in frequency as well as amplitude on relevant specimen of actual welded pipes. More details on such tests can be found in [9, 10, 11, 12, 13]. The specimens were 300 mm long, 10mm thick and 386 mm diameter cylinders (Fig.1).

Thermal cycling was applied on a 70 mm wide sector of inner surface. The cold part of the cycle was obtained by pulverization of a water spray on hot surface and the heating was obtained by infra-red radiations. During the test the INTHERPOL specimen was rotated in order to expose the tested sector alternatively to the spray and to the infrared radiations (Fig.1). For thermal fatigue test monitoring, five thermocouples were laid on the studied internal sector along z axis (Fig.1a). During the tests, the external hot units maintained the external surface of the specimen at 230-240°C. The internal sector was 230-240°C heated by the infrared radiation then cooled to 100-120°C, by the cold spray. The recorded temperatures were then used as input data to perform thermo-mechanical Finite Element simulations. For more details, see [12]. Tests were carried out with temperature amplitudes ΔT of 120°C between the inner and external surfaces. The cycle's duration was ranging from 5 to 8 seconds. Two kind of surface roughness were analyzed: raw machined surface and smooth brushed surface.

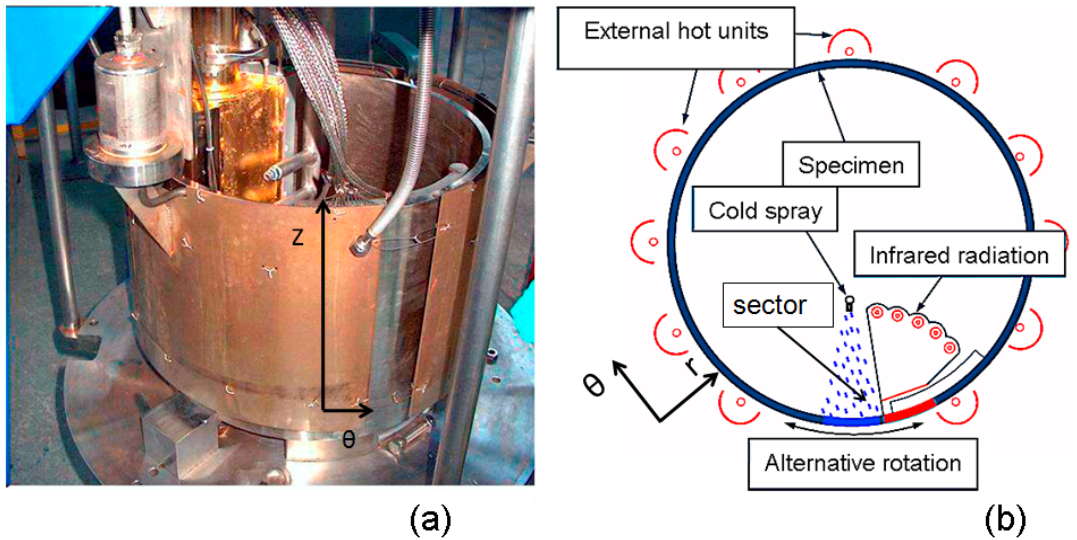

**Fig.1**. (a) Tubular specimen, z is the axial axis, (b) Thermal fatigue test of a sector of the tubular specimen.

2.2 Characterization of 304L stainless steel

*2.2.1 Chemical composition*
The stainless steel used for this study is an AISI 304L quenched from 1040 °C. Its chemical composition is given in Table.1.

| Element | C | Mn | Si | S | P | Ni | Cr | Cu | N2 |
|---|---|---|---|---|---|---|---|---|---|
| Weight% | 0.014 | 1.62 | 0.34 | 0.005 | 0.015 | 10.4 | 18.3 | 0.03 | 0.07 |

Table.1. Chemical composition of the 304L stainless steel used for INTHERPOL experiments (weight %).

This composition and heat treatment were selected to avoid the presence of delta ferrite allowing crack initiation at delta ferrite / matrix interface and secondary cyclic hardening due to martensitic transformation. According to literature [14, 15], these phenomena add more complexity to the micro-structural features and cyclic behaviour of the Face-Centered Cubic (FCC) alloy considered here (Stacking Fault Energy =29mJ/m2).

*2.2.2 Mechanical characterizations*

Mechanical characterizations were performed thanks to cylinder specimens (18 mm gauge length, 8 mm diameter) cut out from the initial material. Tensile tests, performed on specimens cut out in different directions of the initial material sheet, showed an isotropic mechanical behaviour (elastic and plastic), in agreement with a negligible crystallographic texture measured by X ray. The main results are given on Table.2.

| Young's modulus | Poisson ratio | Yield strength 0.2% | Ultimate strength | Ductility (elongation) |
| --- | --- | --- | --- | --- |
| 193 GPa | 0.29 | 250 MPa | 594 MPa | 59 % |

Table.2..Tensile properties of the 304L steel used for INTHERPOL experiments.

*2.2.3 Characterization of the internal surface machining by SEM observations*

The surface profile characteristics of the internal surfaces of two tubular specimens are given on Fig.2a and Fig.2b. The inner surface profile of the raw specimens exhibited up to 50µm deep machining grooves (i.e. about one grain size). Raw surface scratches being more crucial for cracks formation than brushed surface, only the results on raw surface will be presented. Observations by Scanning Electron Microscopy (SEM) and by Electronic Back Scattering Diffraction (EBSD) techniques revealed that the mean grain size is 52 µm and most grains showed annealing twins (Fig.2c). One should note that the austenite phase being rather stable (high Ni content), no delta ferrite was observed, neither on micrographs nor on EBSD maps. According to EBSD technique and X-rays micro-analysis performed by Energy-dispersive X ray spectroscopy, ferritic phase was found less than 3% in volume.

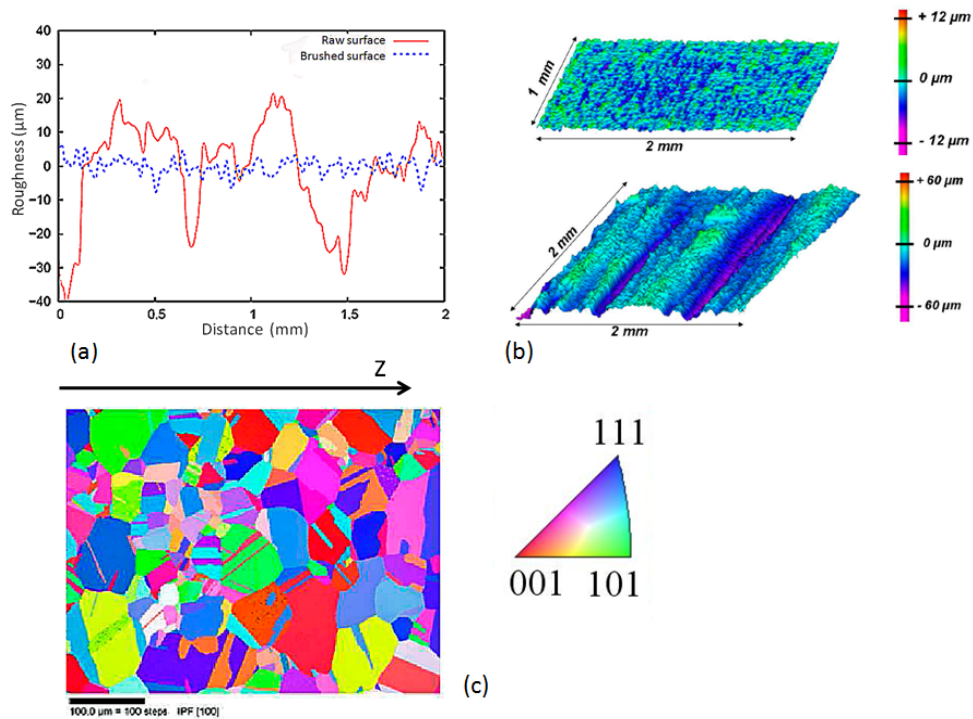

**Fig.2.** (a) surface profiles characteristics, (b) characterisation of the two studied surfaces (brushed and raw) with an Alicona 3D scanning optical microscope, (c) grain orientation map of 304L austenitic steel obtained thanks to EBSD (each colour corresponds to a crystallographic orientation (z is the axial axis of the tubular specimen).

*2.2.4 Internal stress measurement*
J.L. Lebrun (ENSAM Laboratory) performed residual stresses measurements (in axial and tangential directions) at the inner surface of INTHERPOL tubular model. These measurements, performed in the initial state and after thermal fatigue and cooling, revealed a strong relationship between surface finish and residual stresses. Furthermore, the measurements showed that these residual stresses strongly evolved during thermal cycling [16]. This strong evolution was favoured by the difference of mean temperature between the (cooled) fatigue test sector and the main part of the tube, inducing biaxial positive mean strains in the test zone, which are superimposed to cyclic and residual strains. Therefore, cyclic plasticity and /or creep led to subsequent cyclic ratcheting of the test zone. A partial relaxation of the corresponding mean tensile stress (see simulation Fig. 10) and similar evolution of residual stresses were expected. After final cooling, the residual stresses in the test sector reached -200 MPa for the axial direction and -400 MPa for the tangential direction. X ray diffraction peaks width gave some information on the hardening state. These measurements showed a weak and negligible evolution of the hardening state of the tubular specimen after cycling test.

*2.2.5 Dislocation microstructures*
Transmission Electron Microscope (TEM) observations, performed on INTHERPOL specimens, revealed a similar strong hardening gradient (named pre-hardening in the following) on both "raw" and "brushed" surfaces. Only the results on "raw" surface before and after thermal tests are presented.

A very thin partially transformed zone $\gamma \to \alpha'$ martensitic phase (body-centered cubic structure) was observed at less than 2 µm of the surface. Beneath the inner surface, the dislocation density reached $10^{14}$ m$^{-2}$ at a 250-300 µm depth. Some thin twins, due to pre-straining, were observed at the vicinity of the surface. In material, the dislocation density measurement gave $9.10^{12}$ m$^{-2}$.

Fig.3 shows some TEM observations of the austenitic microstructure before and after fatigue tests. After cycling and cooling, we observed some dislocation arrangements (cells formation) but the dislocation substructure near the surface remained dense and entangled. We observed that the dislocation density variations before and after cycling were weak. The low stacking fault energy of the stainless steel alloys (compared to Cu or Al alloys), leads to a rather planar dislocation slip and to an uneasy reorganisation of pre-hardened dislocation substructure. This typical feature of stainless steel alloys is responsible for both strong memory and Bauschinger effects [14], [17], [18].

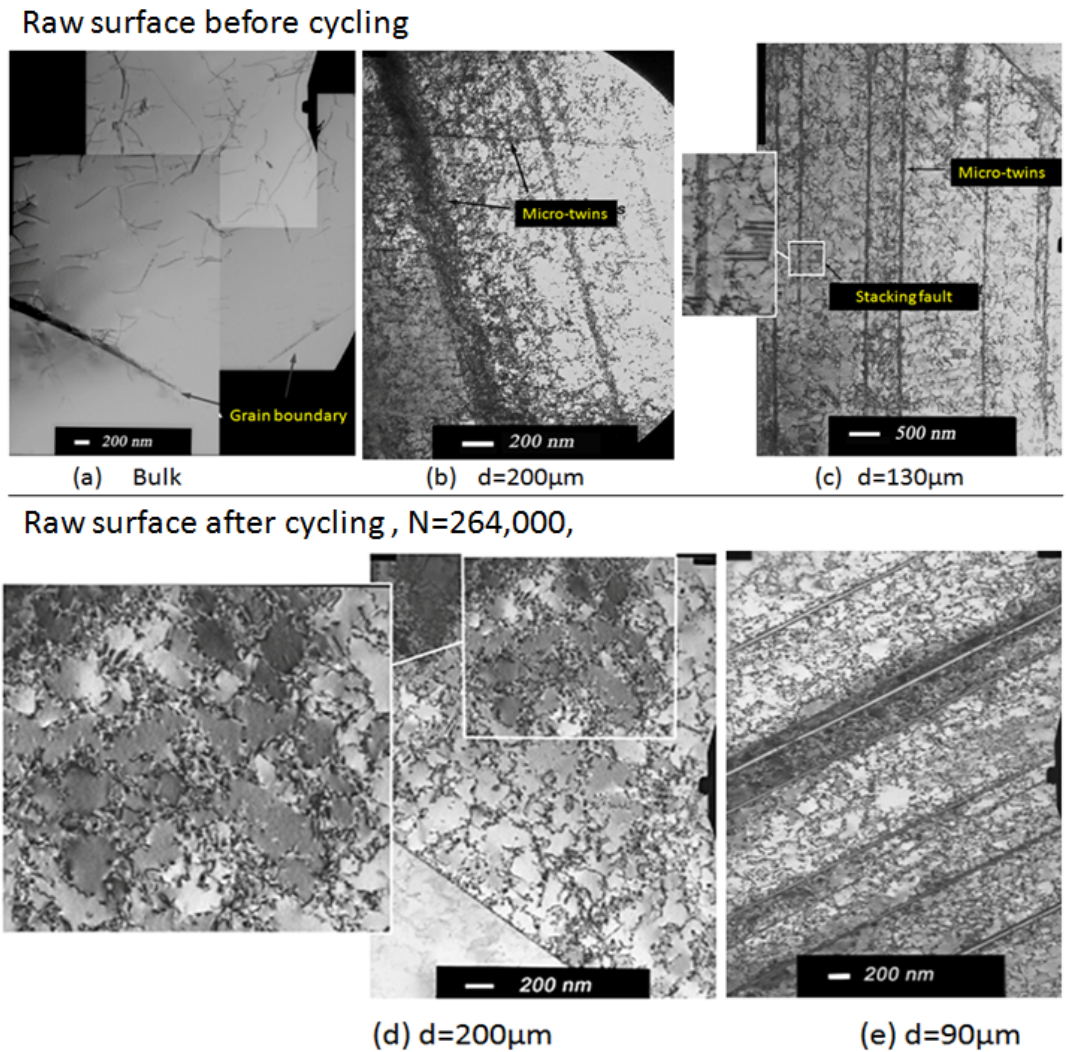

**Fig.3**: TEM observations for different distances from the raw surface. Initial state: (a) weak dislocation density in the material (b) d = 200 µm, twinning, (c) d = 130µm, high dislocation density, twinning and stacking fault. Final state: (d) random dislocations arrangement, (e) high dislocation density and some cell arrangement.

*2.2.6 Micro-hardening tests*

The pre-hardening gradient function of the distance to the surface was characterized by micro-hardness tests (0.05kg) at the initial state and after fatigue tests (Fig.4). These tests showed a strong pre-hardening gradient located on a 250-300 µm layer. It should be noted that this gradient stayed almost unchanged after 265,000 cycles of thermal fatigue. This latter result agrees with TEM observations: the dislocation density is not sensitive to cycling.

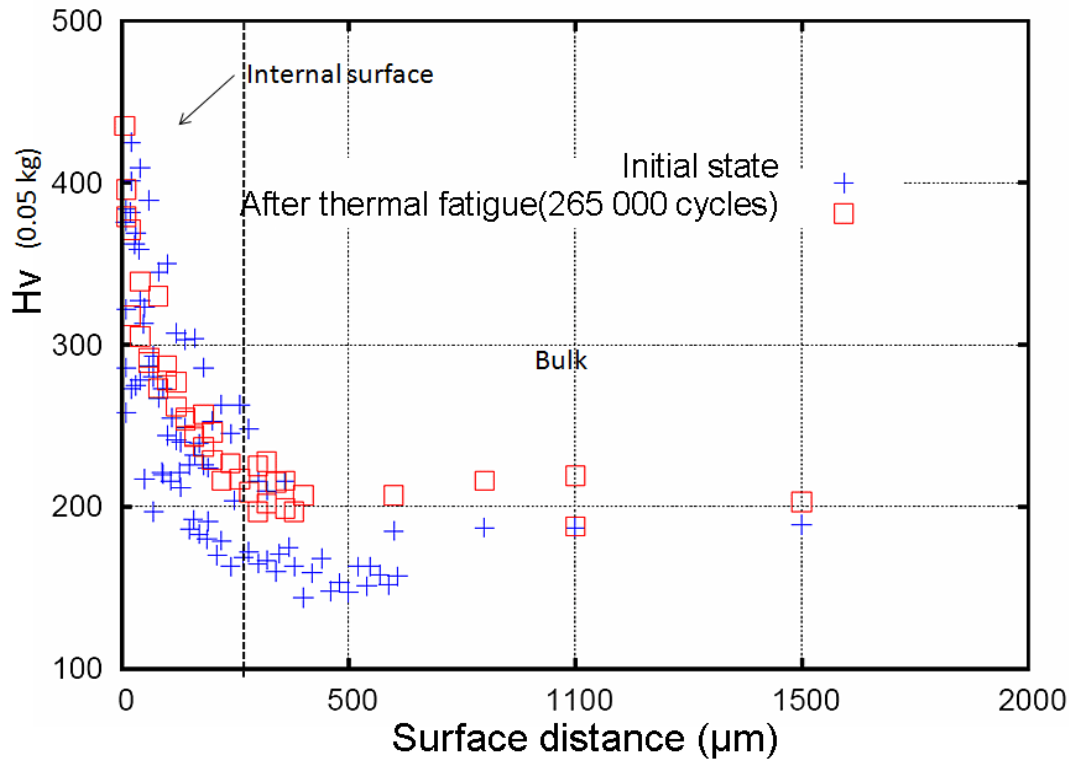

**Fig.4:** Experimental curves of hardness/surface distance before and after thermal cycling.

Pre-hardening gradient corresponds to a gradient of dislocation densities and sub-structure, which is not waived away by thermal fatigue tests. This memory effect, induced by the low stacking fault energy and rather planar slip of stainless steel alloys, has a significant part in the mechanical behaviour of the austenitic stainless steel. This hardening memory effect will have to be introduced in the constitutive equations of 304L, in order to get a good description of mechanical stress and strain fields at the inner surface of the tube, where crack initiation takes place.
Table.3 compares the initial microstructure features and micro-hardness measurements

| Depth(µm) | Microstructure | | Micro-hardness (0.05kg) | |
|---|---|---|---|---|
| Bulk | Raw | Brushed | Raw | Brushed |
| ~ 300 | Weak dislocation density | | 180 | |
| ~~259 | Large dislocation density | | 200 | |
| ~ 200 | Large dislocation density+stacking fault | | 225 | |
| ~ 150 | Large micro-twin density | Large micro-twin density and large dislocation densitu | 250 | |
| ~2 | Large micro-twin | Large dislocation | >400 | 300 |

| | density+nano-grains | density+micro-twins | | |
|---|---|---|---|---|

Table.3. Characterization surfaces before cycling of the raw and brushed.

*2.2.7 Fatigue tests*

Fatigue tests were performed by uniaxial loading on cylindrical specimens (8 mm diameter and 18 mm gauge length). The strain rates were $4\ 10^{-3}\ s^{-1}$ and $2\ 10^{-3}\ s^{-1}$ respectively. For some specimens, the pre-hardening state representative from surface condition was achieved by a 13.6% tensile test prior to cyclic loading. This pre-hardening magnitude was chosen in order to reproduce both the micro-hardness and dislocation density of the material layer induced by machining.

Either without or with pre-straining, the fatigue life curves ($\Delta\sigma/2$-N) showed classical strengthening, followed by softening leading to stabilization or to second hardening for large strain amplitudes [19, 4]. The stress level was higher when the imposed strain range was large (Fig.5a, Fig.5c). For half life fatigue (durability domain) and whatever the applied strain level, the σ-ε curves presented a Bauschinger's effect (Fig.5b, Fig.5d).

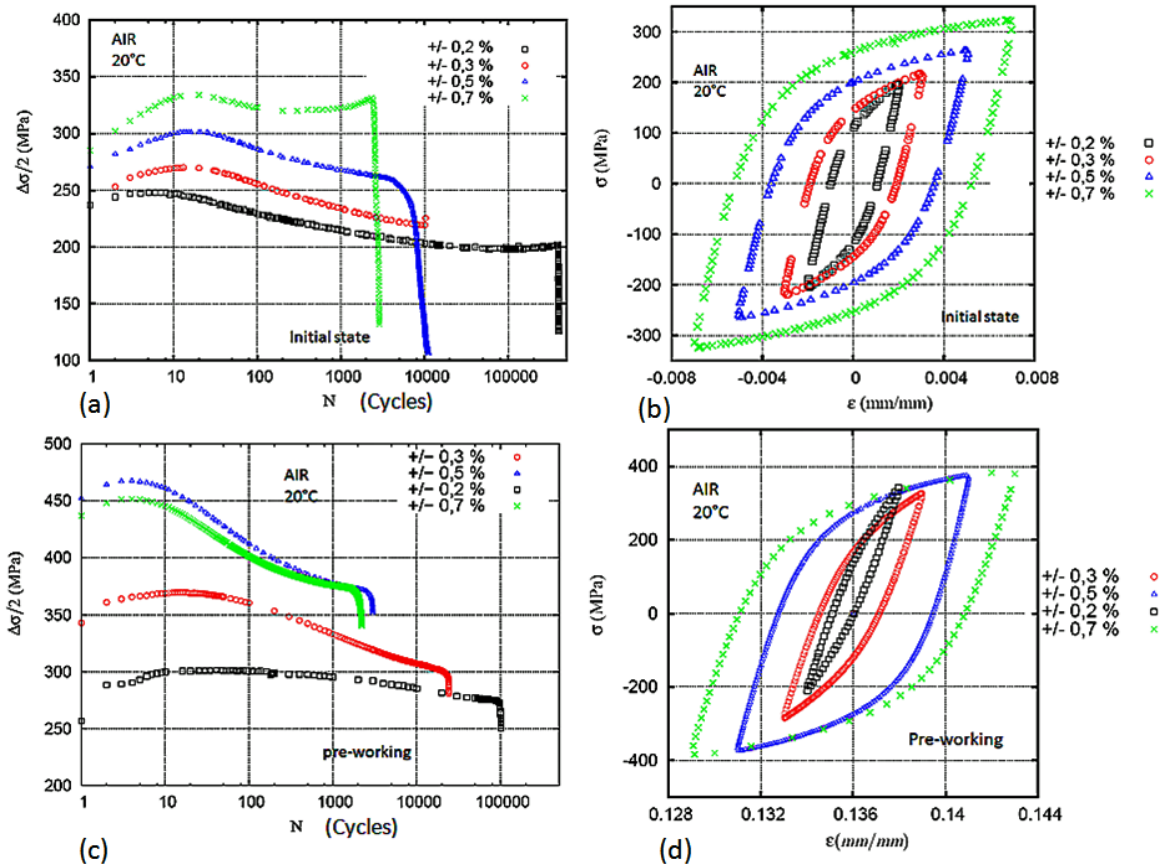

**Fig.5**: Fatigue tests curves ($\Delta\sigma/2$-N) and (σ-ε) curves at room temperature (strain rate of $4\ 10^{-3}\ s^{-1}$). (a), (c) without pre-straining, (b), (c) with 13.6% pre-straining.

Pre-straining in tension (Fig.5c) induced an average tensile stress which increased the stress amplitude, decreased the fatigue life for small strain amplitude (±0.2% and ±0.3%) and led to non symmetric cycles and large Bauschinger's effect (Fig.5d).

It should be noted that, in air, for ±0.3% strain amplitude and at 300 °C, the fatigue life of rough specimens was 4 times lower than the smooth specimens fatigue life. This factor life reduction slightly depends on the strain amplitude. These results are of the same order of magnitude than the indications provided by the report of the Argonne National Laboratory [20]

## 3 Macroscopic FE Modelling

As described in the experimental part, the internal and external surfaces of the cylinders were instrumented with thermocouples. Strain gauges could not be used in the inner thermally cycled surface of the tube without introducing a modification of local thermal transfer conditions and a possible initiation of fatigue cracks. Consequently, the strain and stress states of the specimen during thermal cycling were only available through numerical simulation, and could only be compared to strain gauge measurements on the external surface of the specimens. The stabilized cyclic curve was simulated using the Finite Element software Aster® developed by EDF.

A thermo-mechanical simulation of the whole fatigue tests was performed and the components of the stress and strain tensors due to thermal cycling were determined. The macroscopic modelling rests on a simplified approach of Chaboche's model [6] which can capture both the cyclic thermo-mechanical behaviour and the pre-strain memory effect of the inner surface material. This model was extended to creep, restoration and ratcheting by Chaboche and Nouailhas [21, 22, 6].

Considering the low temperatures of the tests, compared to the recovery and recrystallization temperatures of 304L, the recovery phenomenon was neglected. The pre-strain memory effect, which corresponded to the work-hardening gradient was assumed time independent during thermal fatigue tests. The total strain was the sum of elastic and inelastic (visco-plastic) strains.

### 3.1 Constitutive equations
The yield surface is given by:
$$J_2(\tilde{S} - \tilde{X}) - k - R = 0 \qquad (1)$$
With:
$$J_2(\tilde{S} - \tilde{X}) = \sqrt{3/2(\tilde{S} - \tilde{X}):(\tilde{S} - \tilde{X})} \qquad (2)$$
$\tilde{S}$ is the deviatoric stress tensor, $\tilde{X}$ the sum of two kinematic hardening tensors (showing two different evolution rates) and R the isotropic hardening.

In this model, the saturation effect related to viscosity is described by an exponential function introduced in the visco-plastic potential function.
$$\Omega = \frac{K_0}{(n+1)} exp\left(\left\langle \frac{\sigma_v}{K_0} \right\rangle^{n+1}\right) \qquad (3)$$
with $\langle x \rangle = 0$ for $x \leq 0$

where $\sigma_v$ is the viscosity stress :

$$\sigma_v = J_2(\tilde{S} - \tilde{X}) - k \qquad (4)$$

The visco-plastic strain tensor is derived from the normality law:

$$\dot{\tilde{\varepsilon}}^p = \left(\frac{\partial \Omega}{\partial \tilde{\sigma}}\right) = \frac{3}{2} \dot{p} \frac{\tilde{S} - \tilde{X}}{J_2(\tilde{S} - \tilde{X})} \qquad (5)$$

where $\dot{p}$ is the equivalent (cumulated) strain rate :

$$\dot{p} = \left(\frac{\sigma_v}{K_0}\right)^n \qquad (6)$$

In this work, the memory effect is introduced through the dependence of the asymptotic value $Q$ of the non linear isotropic hardening $R$ from previous straining (more details can be found in Chaboche [21, 6] and Nouailhas, [22])

$$\dot{R} = b(Q - R)\dot{p} \qquad (7)$$

with $Q = Q_0 + (Q_M - Q_0)(1 - \exp-(2\mu q))$

where $\mu$ and $b$ are parameters and $q$ is a variable.

The previous maximal strain is taken into account via the variable $q$ which controls, in the inelastic strain space, the shape of a threshold surface corresponding to the maximum previous strain. As long as the strain loading remains inside this surface, the hardening parameter $Q$ remains constant (memory effect). As soon as a new maximum strain $q$ is reached, this surface evolves and a new value of $Q$ is computed from above equation (7).

$Q_M$, $Q_0$ and $\mu$ are parameters correlated to the memory effect.

The kinematic hardening is composed of two non linear terms:

$$\tilde{X} = \tilde{X}_1 + \tilde{X}_2 \qquad (8)$$

$$\dot{\tilde{X}}_i = \frac{2}{3} C_i \dot{\tilde{\varepsilon}}_p - \gamma_i \tilde{X}_i \dot{p} \qquad (9)$$

and

$$\gamma_i = \gamma_i^0 [a_\infty + (1 - a_\infty \exp-(bp))] \qquad (10)$$

where ($i = 1,2$) and $\gamma_i$ are material parameters

3.2 Identification of the model parameters

Our main objective is the modelling of the stabilized cycles of the different materials (initial, raw and brushed surfaces) and the determination of the elasto-plastic axial and tangential stress-strain curves. The prevision of the $\Delta\sigma/2$-$N$ fatigue curves needs a more complex model and is not studied in this paper.

The constitutive laws involve fourteen parameters identified on the basis of traction-compression tests at room temperature and at 300°C, for specimens with and without pre-hardening. Viscosity parameters are determined from fatigue tests performed at different strain rates: 4. $10^{-3}$ s$^{-1}$ and 6 $10^{-3}$ s$^{-1}$ at 20°C and 4 $10^{-3}$ s$^{-1}$ and at 2 $10^{-3}$ s$^{-1}$ for 300°C). Fifteen cyclic traction-compression tests (with applied controlled total strain ranging between +/-0.2% and +/-0.7%) were performed on the 304L samples taken from the tubular cylinder. Although the thermo-mechanical loading is biaxial and slightly non proportional, tension torsion tests showed that over-strain hardening effect can be neglected here [13].

Table.4 gives the experimental conditions. The results of parameter identification at 20°C and 300°C, are given on Table.5 and Table.6 respectively.

| T | ± $\Delta\varepsilon_t / 2$ | | | | ± $\Delta\varepsilon_t / 2$ | | | |
|---|---|---|---|---|---|---|---|---|
|   | Initial material | | | | 13.6% pre-strained material | | | |
| 20°C | 0.17% | 0.3% | 0.5% | 0.7% | 0.3% | 0.5% | 0.7% | 0.7% |
| 300°C | 0.2% | 0.3% | 0.5% | | 0.2% | 0.3% | 0.5% | |

Table 4. Experimental tests used for model parameter identification ($\Delta\varepsilon^t$ is the total strain range).

| $E(MPa)$ | $\nu$ | $b(nm)$ | $n(-)$ | $K_0(MPa)$ | K(MPa) | $Q_0(MPa)$ | $Q_M(MPa)$ | |
|---|---|---|---|---|---|---|---|---|
| 194,000 | 0.3 | 11.2 | 23.5 | 11 | 132 | -86 | 396 | |
| $C_1(MPa)$ | $C_2(MPa)$ | $\gamma_1^0 (s^{-1})$ | $\gamma_2^0 (s^{-1})$ | $a_\infty (-)$ | $\gamma_2^0 (s^{-1})$ | $a_\infty (-)$ | $\eta$ | $\mu(-)$ |
| 172,840 | 12,613 | 2,729 | 93 | 0.397 | 93 | 0.397 | 0.135 | 8 |

Table.5. Material parameters identified on stabilized traction-compression cycles at 20 °C.

Only E, b, n, k, $Q_0$, $Q_M$ are temperature dependant. Other parameters are unchanged.

| $E(MPa)$ | $b(nm)$ | $n(-)$ | $k(MPa)$ | $Q_0(MPa)$ | $Q_M(MPa)$ |
|---|---|---|---|---|---|
| 170,000 | 2.4 | 74 | 71 | -156 | 1,110 |

Table.6. Material parameters identified on stabilized traction-compression cycles at 300°C.

We assumed a linear variation of E, b, n, k, $Q_0$, $Q_M$ parameters between 20°C and 300°C, in order to determine the actual temperature of INTHERPOL simulation.

Fig.6 and Fig.7 show the numerical and experimental stabilized traction-compression cycles, for different applied strains at 20°C and at 300°C, without and with pre-hardening. The pre-strained stabilized cycles were obtained after a 13.6% tensile test followed by 10 cycles.

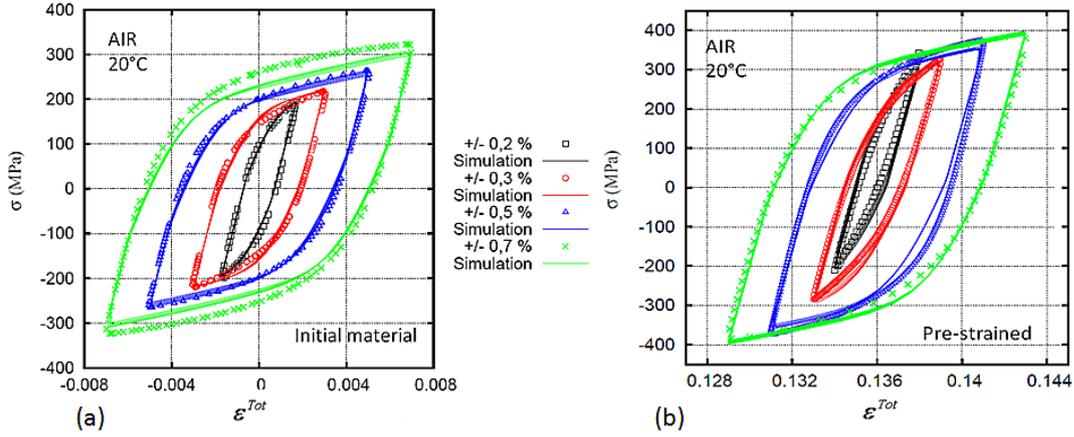

**Fig.6**: Comparison of experimental and simulated stabilized traction compression cycle (20°C, $\dot{\varepsilon}$=4 $10^{-3}$ $s^{-1}$), (a) without pre-straining (b) with 13.6% pre-straining.

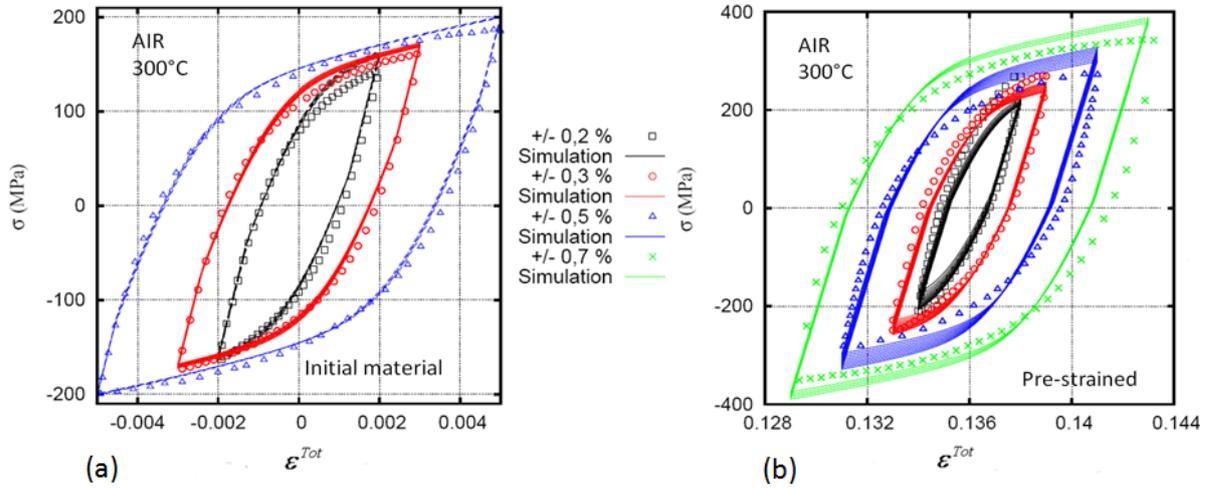

**Fig.7**. Comparison of experimental and simulated stabilized traction compression cycle (300°C, $\dot{\varepsilon}$=4 $10^{-3}$ $s^{-1}$).(a) without pre-straining (b) with a 13.6% pre-straining.

The very good agreement between experimental (dotted lines) and computed cycles (solid lines) at 20°C and at 300°C, shows the ability of the Elastic-Visco-Plastic (EVP) Chaboche's model to describe the strong memory effect (pre-straining effect) observed on the 304L steel. Nevertheless, due to the Armstrong Frederiks kinematic hardening law [23] used in this study, such a model cannot be easily extended to both monotonic and cyclic loadings as shown by the numerical tensile curve obtained. The numerical hardening slope is steeper than the experimental one [12].

3.3 Meshing, boundary conditions and numerical results
The simulation of the thermo-mechanical cycles of the INTHERPOL structure was performed by Curtit and Stephan [3, 4], with Code‑ASTER® developed by EDF.

*Meshing:* the meshing was composed of 11,512 quadratic (20 nodes) elements. In the axial direction, the size of the elements grew with the cylinder thickness. The thinner elements were on the internal surface. The meshing was divided in 50 sectors. A refined meshing was used for the sector submitted to cyclic cooling and heating, as presented on Fig.8. For a given sector, the duration of the cycle was 5s, with a time step of 0.1 s

*Boundary conditions*: The boundary conditions were applied at three points named P1, P2, P3 of the bottom of the cylinder structure (isostatic conditions). The imposed displacements were:

P1 : $u_r = 0, u_\theta = 0, u_z = 0$

P2 : $u_r = 0, u_z = 0$

P3 : $u_z = 0$

*Loading path*: on the external surface the imposed temperature was $T_e=180°C$. On the inner surface, the initial temperature was $T_i=20°C$. The imposed temperature flow across the section is $12 kW/m^2$.

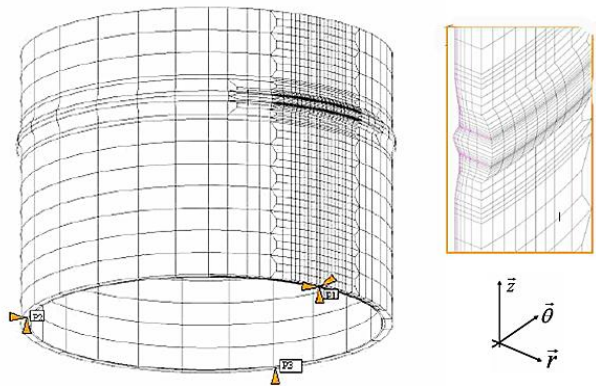

**Fig.8**. Meshing of the INTHERPOL specimen. In the right insight, the shape of the mesh through the thickness (the weld effect is not studied in this paper).

Two simulations were successively undertaken: a thermo-elastic simulation and then an elasto-visco-plastic simulation. The thermo-elastic simulation gave the stabilized temperature field on the internal surface of a sector. To avoid time consuming computation, simulation was stopped as soon as the temperature cycle in the first layer of the elements of the internal surface was stabilized. All details of the simulation and results were published [9, 10]. The thermo-elastic model was validated by comparison of simulated and measured strains at the external surface as well as the simulated and measured temperatures on both surfaces. The thermal fields calculated during the last stabilized cycle were then used as input data for the mechanical field simulation. The $\Delta T$ stabilized difference temperature between the points of the internal surface of the concerned sector and the specimen was about 80°C as shown on Fig.9.

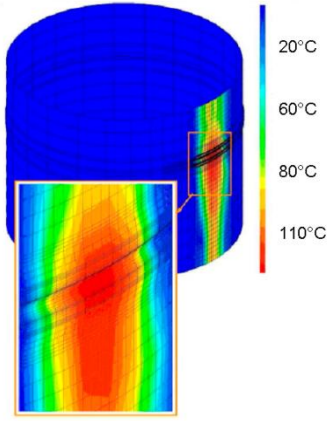

**Fig.9**. Computed temperature difference $\Delta T$ (internal surface) between the concerned sector and the other sectors of the specimen. These sectors are printed in blue colour.

The mechanical fields were computed for the whole structure, thanks to an elasto-visco-plastic simulation, for a cycle of 5s with a 0.1s time step. For a given temperature, the parameters were interpolated from Table 4 and Table 5. In this way, the pre- hardening and the internal stresses were taken into account.

Initial residual stresses were computed via a thermo-mechanical simulation corresponding to a simulated thermal shock (80°C to 20°C) applied to the internal surface. The computed values of the residual stresses were found close to the experimental initial values obtained by X ray. The simulated initial residual axial and tangential stresses $\sigma_{zz}^{i} = 200\ MPa$, $\sigma_{\theta\theta}^{i} = 400\ MPa$ were in good agreement with the experimental initial residual stresses $\tilde{\sigma}^{i}$ measured by X ray (ENSAM-Angers).

The thermal cyclic loading is applied to the whole structure. 40 cycles are simulated in order to obtain the mechanical stabilized cycle.

At each step of the simulation, the total elasto-plastic strain field $\tilde{\varepsilon}^{T}$ was obtained by: $\tilde{\varepsilon}^{T} = \tilde{\varepsilon}^{sim} - \tilde{\varepsilon}^{th}$, where $\tilde{\varepsilon}^{sim}$ was the computed strain field and $\tilde{\varepsilon}^{th}$ ($\varepsilon_{rr}^{th} = \varepsilon_{\theta\theta}^{th} = \varepsilon_{zz}^{th} = \alpha(T)\Delta T$) was the thermal dilation field. $\alpha(T)$ was the dilatation coefficient corresponding to the local temperature.

The results concerning the elasto-plastic stress-strain fields $\sigma_{zz}^{T}/\varepsilon_{zz}^{T}$ and $\sigma_{\theta\theta}^{T}/\varepsilon_{\theta\theta}^{T}$ are given on Fig 10.

Since only a sector of the tube being cyclically cooled, cyclic shakedown, ratcheting and structural effects took place. Both stress and strain evolved during cycling, leading to a quasi-symmetric loading in axial direction and to a non symmetric positive loading in tangential direction (positive average stress in the cooled sector).

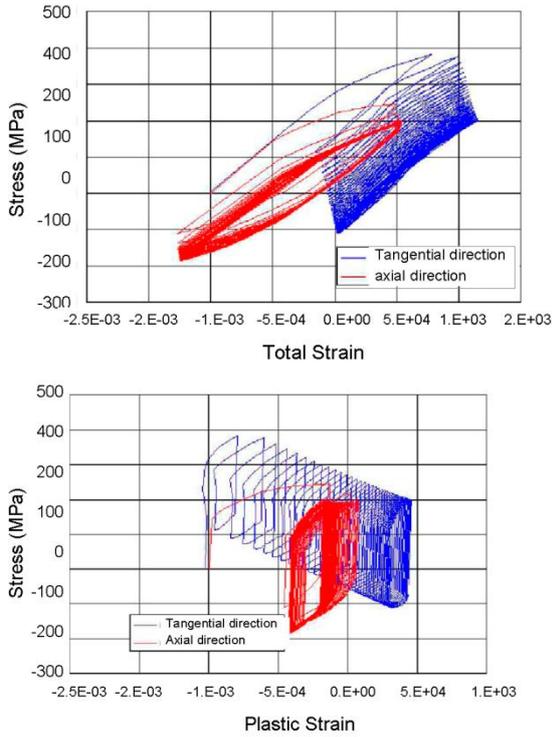

**Fig.10**. Evolution of the elasto-plastic stress-strain parallel to the directions $\vec{z}$ and $\vec{\theta}$ (20°C, $\dot{\varepsilon}$=4 10$^{-3}$ s$^{-1}$).

The minimal, maximal and average stress values of the internal surface are given on Fig.11. These curves show that the use of a fatigue criterion, taking into account the mean (or hydrostatic) stress or strain [24, 25, 26, 27], should be applied using the actual mechanical conditions at the inner surface of the tube, that occur after stabilization of stress and strain fields.

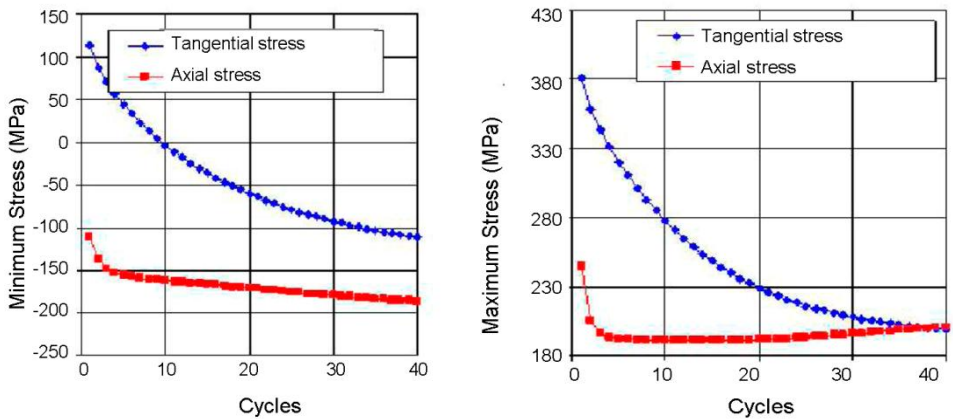

**Fig.11**. Evolution of the minimum and maximum principal stresses at a central point of the internal surface.

Although the stabilized state is not fully reached after 40 simulated cycles, the corresponding decrease of the computed average stresses $\langle \sigma_\theta \rangle$ and $\langle \sigma_z \rangle$ is of the same order of magnitude as the measured ones. The comparison with X ray measurements is given in Table.7.

|  | Experimental decreasing (X ray) | Numerical decreasing |
|---|---|---|
| $\langle \sigma_\theta \rangle$ | -400 MPa | -322 MPa |
| $\langle \sigma_z \rangle$ | -300 MPa | -192 MPa |

**Table.7.** Comparison of the decreasing of the average principal stresses $\langle \sigma_\theta \rangle$ and $\langle \sigma_z \rangle$.

The computed ratcheting (Fig.10) is induced by the difference of mean temperature between the test section (cyclically cooled and heated) and the constantly heated other sectors of the pipe. The temperature of the test section being lower than the other part of the tube, the negative difference of mean thermal expansion has to be balanced by a mean positive mechanical strain on both theta and z axis. The thermal cycling compels the test section to reach a visco-plastic state (at both ends of each cycle) and ratcheting take place. It tends to a progressive plastic elongation of the test section. Consequently, a progressively relaxation of the mean positive stress occurs. After complete relaxation through ratcheting, the test section is submitted to a symmetric stress strain cyclic loading. This evolution can be seen on Fig.11, where relaxation is noticeable, but not completely reached. After final cooling of the whole pipe specimen, the test section was plastically strained in tension during the test (positive biaxial mean strain induced by ratcheting) is submitted to negative biaxial residual stresses (Table.7)

The computed internal strain field evolution is given on Fig.12. The two components presented a cyclical non symmetrical behaviour.

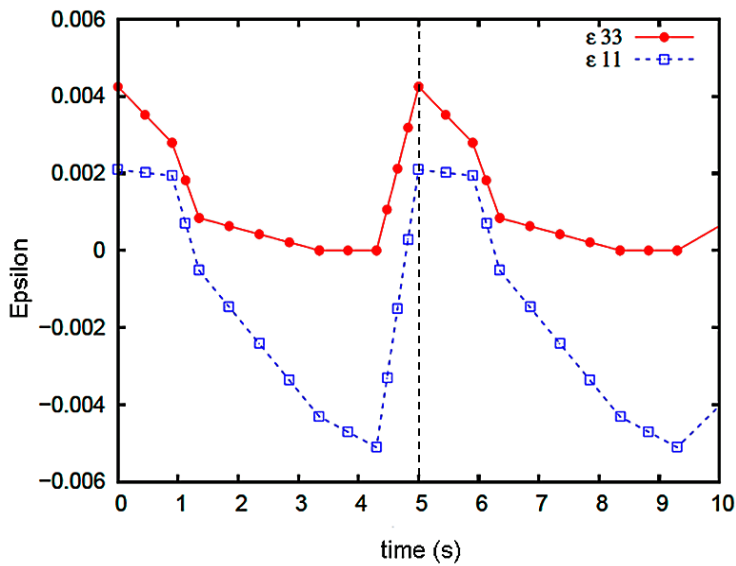

**Fig.12**. Evolution of the internal strain field. Axial strain: $\varepsilon_{33}$, tangential strain: $\varepsilon_{11}$.

In order to determine, through a polycrystal modelling and damage criteria, the micro-crack initiation within the material microstructure, this internal strain field evolution will be applied, in a further paper, to an elementary volume element of the material.

## 4 Discussion and conclusion

INTHERPOL thermal fatigue mock-up was designed for testing of realistic 304L tubular specimens (material, surface finish, welds) under well defined thermo-mechanical loadings. In order to link the thermo-mechanical loading, the microstructure and surface conditions with the thermal fatigue resistance, a detailed analysis of the local conditions prevailing in the subsurface of the tubular specimen was performed. The microstructure and X ray tests carried out on tubular specimen revealed a work hardening gradient and residual stress on the inner surface due to machining surface features. A simulation of thermal fatigue test of the 304L full size tubular specimen was performed, via a non linear 3D FE modelling, with memory effect related to the strong and cyclically persistent pre-hardening gradient on the internal surface of the rough (machined) specimens. Results show that strong structural plasticity effects due to cyclic ratcheting and shakedown must be taken into account to compute the actual macroscopic stress and strain fields within the tubular specimen. Furthermore, this 3D simulation gives access to the actual cyclic stabilized biaxial and non linear loading of the inner surface of the tube, which are far from the initial loading conditions.

The study of initiation of damage at the inner surface of 304L, based on the determination of the stress and strain fields induced by INTHERPOL thermal fatigue cycling, can lead to a better selection and use of fatigue criteria and a better understanding of safety margins or fatigue problems that may arise in actual heat exchangers (submitted). Furthermore, it gives an access to the surface mechanical boundary conditions, which is a prerequisite for a better understanding of the respective effects of crystallographic microstructure (crystal plasticity and crystallographic orientation), pre-hardening gradients (memory effects) and actual roughness geometry. Its can altogether lead to the choice of more fatigue resistant surface machining.

**Acknowledgements**

This work was financially supported by EDF R&D French Company.